\documentstyle[aps,prl,multicol,epsf]{revtex}

\newcommand{\PRL}{Phys. Rev. Lett.}
\newcommand{\PRB}{Phys. Rev. B}
\newcommand{\figref}[1]{Fig.\ \ref{#1}}

\begin{document}
\draft

\title{Numerical investigation of the thermodynamic limit for ground
states in models with quenched disorder}
\author{A. Alan Middleton}
\address{Department of Physics,
Syracuse University, Syracuse, New York 13244}
\date{\today}

\maketitle

\widetext %%%%%%%%

\begin{abstract}
Numerical ground state calculations are used to study
four models with
quenched disorder in finite samples with free boundary conditions.
Extrapolation to the infinite volume limit indicates that the
configurations in ``windows'' of fixed size converge to a unique
configuration, up to global symmetries.  The scaling of this
convergence is consistent with calculations based on the fractal
dimension of domain walls.  These results provide strong evidence for
the ``two-state'' picture of the low temperature behavior of these
models.  Convergence in three-dimensional
systems can require relatively large windows.
\end{abstract}


\begin{multicols}{2}
\narrowtext

The structure of the thermodynamic set of states of a system in
statistical mechanics is studied formally through the infinite volume
limits of correlation functions \cite{RuelleEtc}.
If a nested sequence of
systems with given Hamiltonian and boundary conditions
has
spin correlation functions that converge in the
infinite volume limit, a thermodynamic state can be
defined.
For example, in a ferromagnet with fixed, positive fields at the
boundary, the single-spin correlation function converges to a positive
value, defining an ``up'' state.
For disordered spin systems, the question of the number of
thermodynamic states is a subtle one
\cite{BinderYoung,FisherHuseStates,NewmanSteinCSD,NewmanSteinPQ,NewmanSteinWindow,MigdalOverlap}.
Whether there are many thermodynamic states in some sense
\cite{Parisi} or a small number of states related by simple global
symmetries \cite{FisherHuseStates} (e.g., two spin-flip related states
in an Ising spin glass) has been
a most controversial point for low-dimensional systems.
Part of this
debate has been over what are the most useful methods for determining
the structure of thermodynamic states, spin overlaps $P(q)$
\cite{Parisi,RBY,Overlap3D} or correlation functions in subsystems
\cite{FisherHuseStates,NewmanSteinPQ} and it is unclear whether Monte
Carlo simulations at finite temperature can be used to study large
enough systems \cite{thermalwindow}.

This letter describes the results of numerical computations which
address the structure of states in disordered systems in the
thermodynamic limit, at zero temperature.
Two two-dimensional
models, an Ising
spin glass and a charge density wave (CDW) model (also referred to here
as an elastic medium model)
and two three-dimensional
models, a CDW 
model and a dimer matching model that is equivalent to
non-intersecting lines in a random medium (similar to
vortex lines in type-II superconductors),
were studied.
The
ground states were computed for a sequence of free boundary conditions
and the configurations in a fixed finite subsystem (or ``window'')
were compared.
This study is a particular instance of the numerical approach
suggested by Newman and Stein \cite{NewmanSteinWindow}, who have presented
detailed arguments that the existence of many states, as in the
Parisi solution \cite{Parisi} of the mean field spin glass, gives rise to
``chaotic size dependence'' \cite{NewmanSteinCSD}.
The principle result derived from the simulations presented here is that
the window configurations converge to a {\em single fixed
configuration with probability one}.
These computations strongly support the picture of a
small number of ground states related by global symmetries,
consistent with the droplet model
\cite{FisherHuseStates,BrayMoorestates}.  The details of
the convergence to a single fixed configuration as the boundary
grows has a scaling behavior which is well-described by a simple
picture of domain walls.

The 2d spin glass model (SG) studied has spins $s_i = \pm 1$
defined at lattice points $i$, with Edwards-Anderson Hamiltonian
\cite{EA}
$H_{\rm SG} = - \sum_{\left<ij\right>} J_{ij} s_i s_j$,
where $J_{ij}$
is chosen independently from a Gaussian distribution
for all nearest neighbor
bonds $\left<ij\right>$.
This model is believed to be
paramagnetic at finite temperature, but is a spin glass
at $T=0$; minimal energy
large scale excitations of size $L$ have an energy $E(L)
\sim L^{\theta_{\rm SG}}$ with $\theta_{\rm SG} \approx -0.27$
\cite{2dspinglass}.  The discretized
CDW or elastic medium model in two dimensions
(E2) studied here
is also equivalent to a
disordered substrate model or vortex lines in two dimensions
pinned by quenched disorder
\cite{superrough,gstate}.
The configurations 
in this model are defined by complete dimer
coverings of a hexagonal lattice, with the Hamiltonian being the sum
over covered dimers $d$ of dimer weights $w_d$,
$H_{E2} = \sum_{d} w_d$,
where the $w_d$ are chosen for each bond from a uniform
distribution.  In mean field replica calculations, matching problems
are found to have replica symmetric solutions \cite{MezardParisi,HoudayerMartin}.
A mapping of the dimer model to a discrete height
representation $h$ can be made \cite{BloteHilhorst}; the variable $h$
corresponds to the scalar phase
displacements in CDW models.  This model is
believed \cite{superrough} to have a finite temperature phase
transition, with the height-height correlations
$\left<h(r)h(0)\right>$ behaving as $\sim \ln(r)$ in the high-$T$
phase and as $\sim \ln^2(r)$ in the low-$T$ phase.  In this model,
$\theta_{\rm E2} = 0$ ($E(L) \sim {\rm const}$.)
The model E2 can be extended to three dimensions in two
distinct ways; both are both studied here.
One extension
is that of dimer covering (matching) on a cubic lattice (M3),
which can be mapped to a set of vortex lines with hard-core repulsion
\cite{gstate,defect3dmatch}.
It has
a Hamiltonian identical to that for E2, with the
covering dimers a subset of the edges in a simple cubic lattice.
The other 3-D model is
the three dimensional CDW or elastic medium
model (E3)
\cite{McNamaraEtal};
in the continuum limit, $\theta_{E3} = 1$ (consistent with numerics in
Ref. \cite{McNamaraEtal}.)  The low temperature phase of the elastic
medium models have been studied using both
\cite{LeDoussalGiamarchiOstlundFerndandez} renormalization
group and replica symmetry breaking techniques,
which are usually, though not exclusively,
interpreted physically as describing systems
with few states or many states, respectively.

These models of disordered systems were studied using polynomial-time
combinatorial optimization algorithms \cite{gstate,combopt,barahona}.
The spin glass was studied on a triangular lattice, using the method
developed by Barahona \cite{barahona}, rather than the string method
which is often used \cite{spinstring}.  The minimum weighted matching
algorithm \cite{MoretShapiro} used for the implementation of
Barahona's algorithm was the algorithm described in Ref.\
\cite{cookrohe}.  Calculations were made for at least $10^3$ samples
of up to $512^2$ spins.  The model E2 can be mapped to a bipartite
matching problem \cite{gstate,MoretShapiro} and was solved using the
algorithm of Ref. \cite{Goldberg} for at least $10^3$ samples of sizes
up to $1024^2$ sites.  The same algorithm was used for model M3, 3-D
matching, with up to $128^3$ sites with at least $10^3$ samples, while the
push-relabel maximum flow algorithm as implemented in Ref.\
\cite{Goldberg} was used to study the 3-D
elastic medium model E3 (up to $64^3$ sites with at least $10^3$ samples).
The algorithms used determine ground states up to global symmetry
transformations. For example,
in the spin glass, unsatisfied bonds (bonds with
$J_{ij} s_i s_j < 0$) are calculated, rather than $s_i$.
Configurations related
by symmetries are considered identical here, so that a ``two-state''
picture for spin glasses naturally appears as a single
state in the computations \cite{degeneracy}.

The effect of system size was studied extensively for free boundary
conditions.  The disorder realizations for each sample $S_L^\alpha$ of
linear size $L$ (with $L^d$ spins or sites) were generated so
that $S_L^\alpha$ was a subsystem of a given infinite volume sample
$\alpha$.  Two finite samples $S_L^\alpha$ and
$S_{L'}^\alpha$ had the same quenched disorder in their intersection.
Each finite sample was centered at an origin $C$, so that a sequence
of samples $S_L^\alpha, S_{L'}^\alpha, S_{L''}^\alpha, \ldots$ with $
L < L' < L'' < \ldots$ gives a nested set of square or cubic samples
centered about $P$.  The $L \rightarrow \infty$ limit could then be
numerically studied for a number of infinite samples $\alpha$.
The free boundary conditions were assumed to be typical for the
models SG and M3.
In the elastic models, free boundary conditions 
give ground states with
lower energy than boundary conditions
that would introduce a uniform strain in the elastic models in
the infinite volume limit; such uniform strain states are not considered
here.

\begin{figure}
\begin{center}
\leavevmode
\epsfxsize=5.5cm
\epsffile{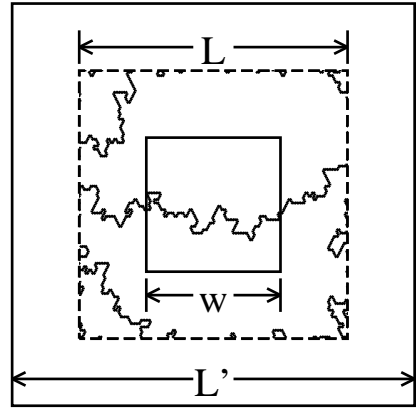}
\end{center}
\caption{ Example of expansion of boundary conditions for the two-dimensional
spin glass (SG).  The ground state for $L'=80$ and $L=80$
subsystems of a single infinite sample are compared.
The solid lines inside the $L=80$ region (dashed box) indicate the
difference (relative domain walls)
in the two ground states in their common area.
The solid box indicates a window of size $w=40$.  In this example, the
expansion of boundary conditions changes the ground state inside of
the window by the introduction of a domain wall that crosses the window.
As can be seen, domain walls exist near the edge of the $L=80$
subsystem; most do not propagate into the middle of the region.}
\label{visual}
\end{figure}

The configuration differences for samples of different sizes $L < L'$
were computed by comparing the exact ground states in the volumes of
size $L^d$ where $S_L^\alpha$ and $S_{L'}^\alpha$ overlapped.
Spin glass ground states in two dimensions were
compared by finding the differences in unsatisfied bonds. 
An example of such a
ground state comparison by bond overlap is shown in
\figref{visual}.
For the
models with dimer matchings (E2 and M3), the configurations are
compared by finding the symmetric difference of the dimer sets in the
common volume.
The natural comparison for the height configurations
for the model E3 is to determine where the {\em gradients}
of the heights in the intersection volume differ.

The primary quantity of interest that was computed was the (sampled)
probability that a change in boundary conditions resulted in {\em any
change} in the ground state configuration in a window of size $w$
centered at $C$.  The probability $P(L', L, w)$ is defined as the
probability that the configuration in the window region changes as the
system size is increased from $L$ to $L'$, that is, that the ground
state configuration for $S_{L'}^\alpha$ differs from that for
$S_{L}^\alpha$ in the volume of size $w$ centered at $C$.
This quantity was estimated by
sampling over a large number of samples $\alpha$
for various $L'$, $L$, and $w$ \cite{statistics}.
This measurement is sensitive to all
gauge invariant spin correlation functions in
the window volume.

\begin{figure}
\begin{center}
\leavevmode
\epsfxsize=6.0cm
\epsffile{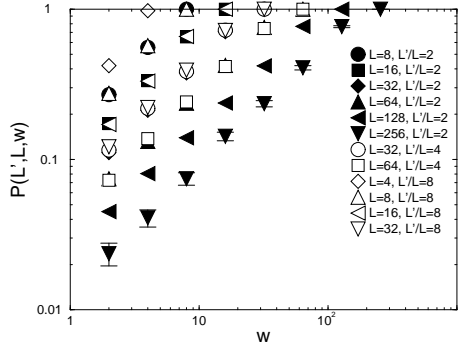}
\end{center}
\caption{A plot of the
probability $P(L',L,w)$ that an expansion in boundary
conditions from $L$ to $L'$ will change the configuration in a
window of size $w$ embedded in the original system of size $L$, for the
two-dimensional spin glass (SG).  Error bars indicate 1$\sigma$ statistical
uncertainties. For large values of $L/w$, $P$ converges to a constant.
For fixed $w$, $P$ decreases as a power law in $L$ (by the even
spacing of the data points at fixed $w$.) Note that for the values of
$L'/L$ shown, $P$ is independent of $L'/L$.}
\label{rawdata}
\end{figure}

A plot of the data for $P(L',L,w)$, as a function of $w$ for
various $L'$ and $L$, is shown in Fig.\ \ref{rawdata} for the
spin glass. Assuming scale invariance, $P$ should be a function of
the two ratios $L'/L$ and $L/w$.
The data is consistent with this hypothesis, for large values
of $w$ and $L$.
For fixed
$L/w$, $P(L',L,w)$ approaches a constant for large $w$ or large $L$.
Note that to within error estimates,
$P(L',L,w)$ is independent of $L'$ for $L'/L=2,4,8$:
the probability of change in a finite window is approximately
{\em independent} of the magnitude of expansion in the boundary, for $L' \ge
2L$ ($P$ does decrease noticeably as $L' \searrow L$.)
In addition, for fixed $w$,
$P(L',L,w)$ decreases approximately as a power law in $L$ (by the even
vertical spacing of the data points for $P \ll 1$.)
The data strongly suggest, by
extrapolation to larger values of $L$ for fixed $w$, that {\em
the probability
of changing the configuration in a window of size w goes to zero for
$L'/L \rightarrow \infty$ as $L \rightarrow \infty$,
implying convergence to a unique
thermodynamic ground state} (up to global symmetries).

\begin{figure}
\begin{center}
\leavevmode
\epsfxsize=8.0cm
\epsffile{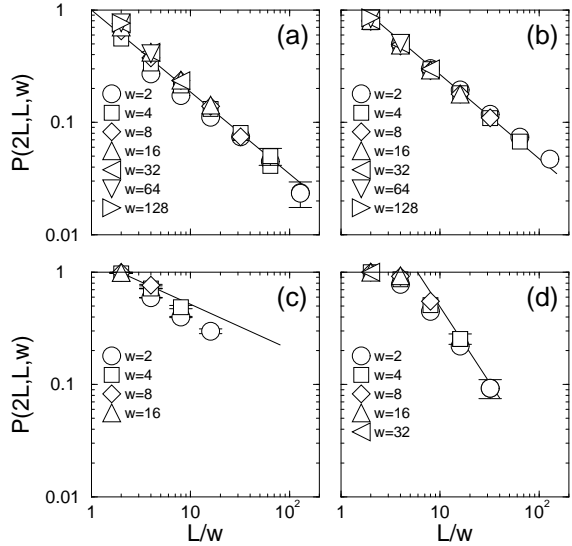}
\end{center}
\caption{ (a) Data for the $d=2$ spin glass (SG) scaled by plotting
$P(L',L,w)$ vs.\ $L/w$.  For clarity, only data for $L'/L=2$ is
shown.  The straight line indicates the slope predicted by $\kappa = d
- d_f = 0.73$.  The data apparently converge to this
form as $w \rightarrow \infty$.  (b) Scaled data for the $d=2$ CDW or elastic
medium model (E2);
the straight line again indicates $\kappa = d - d_f$, with $\kappa = 0.75$.
(c) A scaled
plot of $P(2L,L,w)$ for the 3-D elastic medium (E3); the straight line has
slope $d_f - d = -0.40$, which
approximately parallels the data for fixed $w$.
(d) A scaled plot of $P(2L,L,w)$ for the
3-D dimer matching problem (M3); the straight line has a slope $-\kappa
= -1.35$.  }
\label{scaled}
\end{figure}

The data can be explained by simple
assumptions about the convergence of the
configurations as $L \rightarrow \infty$ and the properties of domain
walls or defect lines.  For the spin glass model,
induced domain walls are lines where the bonds change from satisfied
to unsatisfied or {\em vice versa.}
In models which are represented by a matching (E2 and M3) defects
are also line objects and are composed of bonds where the dimer covering
changes.
In model E3, the induced walls are surfaces where the height
gradient changes.
Defect lines have fractal dimension
$d_f^{\rm SG} = 1.27(1)$ for model SG and $d_f^{\rm E2} = 1.25(1)$ for
model E2 \cite{2dspinglass,gstate,defects2d}.  For the 3-D elastic
medium, a shift in boundary conditions introduces a domain wall of
dimension $d_f^{\rm E3}=2.60(5)$ \cite{McNamaraEtal}, while localized
string defects were computed during the course of this work to have
fractal dimension of $d_f^{\rm M3}=1.65(4)$ in the 3-D matching
model.  If the fractal dimension of the defects is large enough ($d_f
> d/2$) that no more than $O(1)$ defects of size $L$
can co-exist in the volume $L^d$,
the expected number of defects of linear size $L$
introduced upon expansion to size $L'$
is bounded above by a constant.  Whether boundary
changes do induce a number of defects that saturate this bound is less
clear {\em a priori}.  For the models where $\theta \le 0$, finite
changes at the boundary are likely to induce as many defects as
possible, as the large scale defect cost is comparable to the cost of
local changes at the boundary.
The probability
that a line or surface will intersect a window of size $w$ is then the ratio
of the
number of volumes of size $w^d$ that intersect the defect to the number
of areas of size $w^d$ in the area $L^d$, giving the form
\begin{equation}
P(L',L,w) = c(L'/L)\,(L/w)^{-\kappa},
\end{equation}
for large $L/w$,
with $\kappa = d - d_f$ by the
supposition of a single dominant defect and, by these numerical results, the
coefficient function $c(L'/L)$
quickly converges to a constant value for $L'/L \ge 2$.
This form
can be checked by plotting $P$ as a function of $L/w$ and
comparing with a line of slope $d_f - d$, as shown in Fig.\
\ref{scaled}. The match between this prediction and the data in $d=2$
is quite good; a two-parameter fit (varying $c(\infty)$ and $\kappa$)
gives exponents that agree with $\kappa = d - d_f$ to within
$0.05$ for models SG and E2.
Differences of this order are within statistical fluctuations
and apparent finite size effects.
In addition, numerical
study of a number of configurations for three values of $L$
(e.g., $S_L^\alpha$, $S_{L'}^\alpha$, $S_{(L')^2/L}^\alpha$) for the
$d=2$ spin glass
suggests that the location of $L$-scale defects
in a volume is nearly independent
of $L'$, giving more support to the conclusion
that there is convergence to a unique state in these models.

The 3-D results also indicate convergence to a single
state, as $P(L/w) \rightarrow 0$ for
$L/w \rightarrow \infty$. The quantitative fits are also
consistent with a defect picture, but have a larger uncertainty.
For the 3-D elastic medium, the data are consistent
with $P \sim (L/w)^{d_f^{\rm E3} - d}$ for fixed $w$,
as shown in Fig.\ \ref{scaled}(c),
though larger sample sizes would be useful.
For the problem $M3$, the behavior $P \sim (L/w)^{-1.35}$,
with $\kappa = d - d_f$, can be fit to the largest $L/w$ values, though
only over a small range.
Note that $P > 0.9$ for $L/w \le 4$ and $P > 0.5$ for
$L/w=8$.
{\em Under an expansion $L'/L=2$,
the configuration in a window usually changes for $L/w < 8$.}
Such change in small systems mimics
the predictions of a many-states picture.

In summary, the infinite-volume limit for four model disordered
systems was studied numerically by computing ground state
configurations in fixed volumes embedded in systems of successively
larger sizes. Strong evidence was found for convergence to a unique
state (up to global symmetries), even in cases where $\theta \le 0$.
The convergence to a unique state in $d=2$ can be understood in detail
by estimating the chance of a defect wall intersecting a given area
upon a boundary change.
The 3d model results are
more qualitative: while it appears that the system converges to a
unique state, the ratio of scales ($L'/L$, $L/w$) required is larger, so
that systems of size $L > 50$ are needed.
Polynomial ground state algorithms are not available for the 3d spin
glass and this system is not directly addressed here,
but these results suggest that one should be cautious in
interpreting finite temperature Monte Carlo results \cite{RBY}
and ground state calculations \cite{Hartmann3DGS} in small systems.

I would like to thank Daniel Fisher for stimulating
discussions and comments on a draft of this paper.
At the completion of this paper,
I became aware of related results for the spin glass
in $d=2$ by Palassini and Young \cite{PalassiniYoung}.
This work was supported in part by
the National Science
Foundation (DMR-9702242) and by
the Alfred P. Sloan Foundation.

%\end{multicols}

\end{multicols}

\begin{references}
\bibitem{statistics}
The sample index $\alpha$ was a function of $L$ and $L'$, but was
independent of $w$,
so $P(L',L,w)$ is not statistically independent from $P(L',L,w')$.
\end{references}
\end{document}